 \documentclass[12pt, letterpaper]{article}
\usepackage{times}
\usepackage[authoryear,round]{natbib}
\usepackage[]{graphicx}
\usepackage[figuresright]{rotating}
\usepackage{verbatim,color,amssymb}
\usepackage{amsmath}					
\usepackage{array}
\usepackage{lineno}
\usepackage{listings}
\usepackage{rotating}
\usepackage{subfig,subfloat}
\usepackage{multirow}
\usepackage{booktabs}

\newcommand{\rsz}[1]{\textcolor{black}{#1}}

\def\bzero{{\mathbf 0}}

\def\S{{\cal S}}

\def\log{\hbox{log}}

\def\var{\hbox{var}}
\def\cov{\hbox{cov}}

\def\Dir{\hbox{Dirch}}

\def\IG{\hbox{Inv-Ga}}

\def\MVN{\hbox{MVN}}

\def\Normal{\hbox{Normal}}

\def\log{\hbox{log}}

\def\var{\hbox{var}}
\def\cov{\hbox{cov}}

\def\Normal{\hbox{Normal}}

\def\Invwish{\hbox{Inv-Wishart}}

\def\P_25_ICML{{\it Proceedings of the 25th international conference on Machine learning}}

\def\bse{\begin{eqnarray*}}
\def\ese{\end{eqnarray*}}
\def\be{\begin{eqnarray}}
\def\ee{\end{eqnarray}}
\def\bq{\begin{equation}}
\def\eq{\end{equation}}

\def\boldbeta{{\mbox{\boldmath $\beta$}}}

\def\trans{^{\rm T}}

\def\b1e{{\mathbf e}}

\def\bq{{\mathbf q}}

\def\bzero{{\mathbf 0}}

\newcommand{\bbeta}{\mbox{\boldmath $\beta$}}

\newcommand{\uB}       {\mbox{\boldmath$B$}}

\newcommand{\uG}       {\mbox{\boldmath$G$}}

\newcommand{\uM}       {\mbox{\boldmath$M$}}

\newcommand{\uP}       {\mbox{\boldmath$P$}}

\newcommand{\uT}       {\mbox{\boldmath$T$}}

\newcommand{\uU}       {\mbox{\boldmath$U$}}

\newcommand{\uW}       {\mbox{\boldmath$W$}}

\newcommand{\uX}       {\mbox{\boldmath$X$}}

\newcommand{\uZ}       {\mbox{\boldmath$Z$}}

\newcommand{\ubeta}             {\mbox{\boldmath$\beta$}}
\newcommand{\ugamma}            {\mbox{\boldmath$\gamma$}}
\newcommand{\udelta}            {\mbox{\boldmath$\delta$}}

\newcommand{\uiota}             {\mbox{\boldmath$\uiota$}}

\newcommand{\umu}               {\mbox{\boldmath$\mu$}}

\newcommand{\upi}               {\mbox{\boldmath$\pi$}}

\newcommand{\uomega}            {\mbox{\boldmath$\omega$}}

\title{A fully Bayesian semi-parametric scalar-on-function regression (SoFR) with measurement error using instrumental variables.}
\author{Roger S. Zoh, Yuanyuan Luan, Carmen Tekwe}

\begin{document}









\maketitle 

\abstract[Abstract]{Wearable devices such as the ActiGraph are now commonly used in research to monitor or track physical activity. This trend corresponds with the growing need to assess the relationships between physical activity and health outcomes, such as obesity, accurately. The device-based physical activity measures are best treated as functions when assessing their associations with scalar-valued outcomes such as body mass index. Scalar-on-function regression (SoFR) is a suitable regression model in this setting. Most estimation approaches in SoFR involve an assumption that the measurement error in functional covariates is white noise. Violating this assumption can lead to under-estimating model parameters. There are limited approaches to correcting measurement error for frequentist methods and none for Bayesian methods in this area.
We present a fully non-parametric Bayesian measurement error-corrected SoFR model that relaxes all the constraining assumptions often involved with these models. Our estimation relies on an instrumental variable which is allowed to have a time-varying biasing factor, a significant departure from the current approach. Our method is easy to implement, and we demonstrate its finite sample properties in extensive simulations. Finally, we applied our method to data from the National Health and Examination Survey to assess the relationship between wearable device-based measures of physical activity and body mass index in adults in the United States.}



\maketitle


\section{Introduction} \label{sec:intro}
“Large-p-small-n” scenarios occur in multiple regression when the number of covariates $p$ exceeds the sample size $n$. There are two main estimation approaches in such circumstances. The first involves an assumption that no a priori order exists for the covariates and entails estimating the regression parameters for each important covariate while driving regression parameters for unimportant covariates to zero, which effectively is variable selection. \cite{hastie2015statistical} and \cite{van2019shrinkage} reviewed this approach from frequentist and Bayesian viewpoints, respectively. The second approach to large-p–small-n regression involves functional data analysis. The fundamental assumption in functional data analysis is that there exists a functional regression parameter, often with a smooth functional form, that describes the associations between a potentially large number or infinite number of covariates and the response variable \citep{ramsay2006}.

Biomedical researchers increasingly use scalar-on-function regression (SoFR), a kind of functional data analysis, to evaluate the relationships between function-valued covariates and scalar-valued outcomes. In these settings, the functional covariates are observed on a dense or sparse grid. Several authors have reviewed SoFR \cite{ramsay2006,wang2016functional,morris2015functional}. Most prior SoFR models did not include non-functional or scalar-valued covariates. We use a broader definition of SoFR that allows for both function and scalar-valued covariates. Namely, let $\{X(t): t \in [0, 1] \}$, $\uZ$, and $Y$ denote the functional covariate, scalar-valued covariate, the scalar-valued response, respectively, in the following SoFR model: 
\vspace{-.12in}
\begin{equation}
 Y_i = \beta_0 + \int^{1}_{0} \beta(t)X_{i}(t)\,dt + \theta_i \;\; \forall i = 1, \cdots, n,\label{eq:intro1}   
\end{equation}
\vspace{-.02in}
where the error terms $\epsilon_i$ are assumed to be independent and identically distributed; i.e.  $\epsilon_i \stackrel{iid}{\sim} \Normal(0, \sigma^2_{\epsilon})$.
Model~(\ref{eq:intro1}) is estimated within a frequentist framework using pre-selected basis functions (such as B-splines, thin plate splines, polynomial splines, and wavelets) or data-driven basis functions from functional principal components combined with some penalty term to impose smoothness \citep{reiss2017methods}.

 The empirical motivation for our approach is the need to evaluate accurately the association between physical activity patterns and body mass index (BMI). We analyzed physical activity data derived from Actigraphs, wearable devices for monitoring physical activity, and used BMI as an indicator of obesity. Actigraphs and similar devices record data frequently, such as for $60$ second epochs, resulting in high dimensional longitudinal data or functional data with complex heterogeneous covariance structures. Several authors have questioned the accuracy with which such devices measure physical activity \citep{feito2012effects,o2020well,an2017valid}. SoFR can be applied to assess how the function-valued data from wearable devices relate to scalar-valued outcomes such as BMI. However, current approaches to SoFR are limited in their ability to correct for measurement error in imprecisely observed functional covariates with complex error structures.
Most approaches to correcting measurement error in SoFR are based on the assumption, that the function-valued covariates are observed at discrete time points with independent error structures. See \citealp{yao2005functional,cardot2007smoothing,goldsmith2013corrected,morris2015functional,reiss2017methods}. In contrast, \cite{chakraborty2017regression} used a multivariate regression calibration approach while allowing for correlated measurement error. More recently, \cite{tekwe2019instrumental} proposed a measurement error-adjusted SoFR model with an instrumental variable (IV) for identifying the model under the assumption of heteroscedastic measurement error structures.
We estimated model parameters with the generalized method of moments (GMM) and obtained bootstrap confidence intervals for the estimated functional parameter. However, the GMM approach is limited by the assumption of an unbiased IV, which maybe not always be satisfied.

We propose a new method for addressing measurement error in SoFR that differs from prior approaches in several ways. First, we adjust for measurement error with a fully Bayesian semi-parametric method, which allows for an arbitrary measurement error distribution. Our Bayesian method eliminates the need for two-stage estimation involving error-free covariates, while allowing for an automatic, accurate quantification of uncertainty in parameter estimates. Second, although we use a function-valued IV to make the measurement error model identifiable, we remove the assumption of an unbiased IV and also allow the IV to have a functional (time-varying) biasing factor. We provide a way to estimate this additional scaling functional parameter. 

In section~\ref{sec:mod}, we specify our model, the priors, and assumptions.
We describe an extensive simulation study in ~\ref{sec:simeval} to evaluate the performance of our approach. Then in section~\ref{sec:application}, we report on applying our method to data from the National Health and Examination Survey (NHANES) in assessing the association between wearable device-based measures of physical activity intensity and BMI in adults in the U.S. In section~\ref{sec:conclu}, we discuss some potential extensions of our method and offer some concluding remarks.

\section{Model } \label{sec:mod}

\subsection{Model Specification}
Suppose $\left\{\uW_i, Y_{i} \right\}$ is the data pair for each individual $i =1, \cdots, n$, where $\uW_{i} = \{W_{i}(t), t \in [0, 1] \}$. Without lost of generality, we assume time is scaled between $0$ and $1$. The SoFR model with functional covariates subject to measurement error is
\be
Y_{i} &=& \alpha_0 + \boldbeta_{z}\trans\uZ_{i} + \int^{1}_{0}\ubeta(t)X_{i}(t)\,dt + \epsilon_i  \label{eq:modeq1} \\
W_{i}(t) &= & X_{i}(t) + U_{i}(t) \label{eq:modeq2},
\ee
which is similar to our previous model \citep{tekwe2019instrumental}. In model~\ref{eq:modeq1}, $\uX = \{X(t), t \in [0, 1] \}$ is the true but unobserved (latent) functional covariate for which we would like to estimate objectively its association $\ubeta(t)$ with response $Y$. However, we only observe $\uW$, a contaminated (error prone) but unbiased proxy of $\uX$. Also, $\epsilon_i$ is the error for the response model and $U_{i}(t)$ the error process for the measurement error part of the model. Without further assumptions on the error structure $\uU$ or the existence of replicate proxies of the functional $\uX(t)$, it is virtually impossible to estimate model~\ref{eq:modeq1} parameters \cite{Carroll2006,yao2005functional,cardot2007smoothing,goldsmith2013corrected,morris2015functional,reiss2017methods}. We propose an SoFR model that includes models \ref{eq:modeq1} and \ref{eq:modeq2} and 
\be
  M_i(t) &\approx& \udelta(t) X_{i}(t) + \omega_{i}(t),  \label{eq:modeq7}
\ee
where $\uM_i = \{M_i(t), t \in [0, 1] \}$ is the observed value of the IV. We assume $\epsilon_i \stackrel{iid}{\sim}\uG$, for some distribution $\uG$ with mean zero and finite second moment;  $E\left\{U_i(s)\right\} = 0$ and $E\left\{U_i(s)U_i(t)\right\} = \sigma_{u,s,t} < \infty$; $E\{\omega_{i}(s)\} = 0$ and $E\{\omega_{i}(t)\omega_{i}(s)\} = \sigma_{\omega,t,s}$; and all model error terms are independent. 
Estimating $\udelta(t)$ is challenging in our Bayesian approach. We assume $\udelta(t)\neq 0$ is known for $t \in \uT$ and describe estimating $\udelta(t)$ in section 2.4. Given the value of $\udelta(t) \neq 0$, estimation is based on the scaled version of  $M^{*}_{i}(t) = M_i(t)/\udelta(t)$, which yields 
\be
M^{*}_{i}(t)&\approx& X_{i}(t) + \omega^{*}_{i}(t). \label{eq:modeq7a}
\ee
The number of parameters in models~\ref{eq:modeq1}, \ref{eq:modeq2}, \ref{eq:modeq7a} can be extremely large, making estimation nearly impossible. Therefore, we reduce dimensions with a basis expansion. Let $K$ be a given set of orthogonal basis functions stacked as $\uB(t) = \left\{b_{1}(t), \cdots, b_{K}(t) \right\}\trans \in \mathbb{R}^{K}$. Applying the basis expansion to the functional covariates, the models in \ref{eq:modeq1}, \ref{eq:modeq2}, and \ref{eq:modeq7} become
\be
Y_{i} &\approx& \alpha_0 + \ubeta_{z}\uZ_{i} + \sum^{K}_{k=1}\gamma_{k}X_{i,k} + \epsilon_i \label{eq:modeq3} \\
 W _{i,k} &\approx & X_{i,k} + U_{i,k}, \mbox{where}\; k = 1, \cdots, K  \label{eq:modeq4} \\
 M_{i,k} &\approx & X_{i,k} + \omega^{*}_{i,k}, \mbox{where}\; k = 1, \cdots, K,   \label{eq:modeq4a}
\ee
where $X_{i,k} = \int^{1}_{0} b_{k}(t)X_{i}(t)dt$, with $b_{k}$ denoting the $k^{th}$ basis function; $\gamma_k = \int^{1}_{0} b_{k}(t)\beta(t)dt$;  $W_{i,k} = \int^{1}_{0} b_{k}(t)W_{i}(t)dt$; $ M_{i,k} = \int^{1}_{0} b_{k}(t)M^{*}_{i}(t)dt$; $U_{i,k} = \int^{1}_{0} b_{k}(t)U_{i}(t)dt$; and  $\omega^{*}_{i,k} = \int^{1}_{0} b_{k}(t)\omega^{*}_{i}(t)dt$.

\subsection{Model Assumptions and Priors} \label{subsec:priors}
To estimate model parameters, we adopt a fully Bayesian approach.
For a stack version of the data and model, we define for each $i =1, \cdots, n$,  $ \uW_{i} = (W_{i,1}, \cdots, W_{i,K})\trans$; $\uX_{i} = (X_{i,1}, \cdots, X_{i,K})\trans$; $\uomega^{*}_{i} = (\omega^{*}_{i,1}, \cdots, \omega^{*}_{i,K})\trans$
; and $\uU_{i} = (U_{i,1}, \cdots,U_{i,K})\trans$.   
To allow for flexibility in the error terms of the model, we assume a truncated Dirichlet Process mixture (tDPM) prior similar to that used by \cite{sarkar2018bayesian}. This prior helps capture departures from symmetric distributions of the error terms. We estimate the parameters conditional on $K$, the number of basis functions. We also use the following conjugate priors that permit straightforward implementation of the Gibbs steps for sampling from our complex joint posterior distributions. \textcolor{black}{In the following prior distribution definition, $\pi_{.,k}$ will denote the weight of the $k^{th}$ component in the mixture distribution.}
$\epsilon_i \stackrel{iid}{\sim} \sum^{K_\epsilon}_{k=1} \pi_{\epsilon,k}\Normal(.|\mu_{\epsilon, k}, \sigma^2_{\epsilon,k})$, where $\sum^{K_\epsilon}_{k=1} \pi_{\epsilon,k} \mu_{\epsilon, k} = 0$, and $(\mu_{\epsilon, k}, \sigma^2_{\epsilon, k}) \sim \mbox{NIG}(\mu_0, \sigma^{2}_0/\nu_0, \alpha_0, \sigma^{2}_0)$, for $k = 1, \cdots, K_\epsilon$; and 
 $\upi_{\epsilon} = (\pi_{\epsilon,1}, \cdots, \pi_{\epsilon,K_\epsilon})\trans \sim \Dir(\alpha_\epsilon/K_\epsilon, \cdots, \alpha_\epsilon/K_\epsilon)$, where $K_{\epsilon}$ denotes the class membership probabilities for the error $\epsilon_i$.
For the measurement error vector, $ \uU_i \stackrel{iid}{\sim}\sum^{K_u}_{k=1} \pi_{u,k}\MVN(.|\umu_{u,k}, \Sigma_{u,k}), $ with $\umu_{u,k} \sim \MVN(.|\umu_{u,0}, \Sigma_{u,0})$, constrained to $\sum^{K_\epsilon}_{k=1}  \pi_{u,k} \umu_{u,k} = \bzero$;  $\Sigma_{u,k} \sim \Invwish(.|\nu_{u,0}, \Psi_{0,u})$; and $\upi_{u} = (\pi_1, \cdots, \pi_{K_u})\trans \sim \Dir(\alpha_u/K_u, \cdots, \alpha_u/K_u)$.
For the error model of the IV, $ \uomega^{*}_i \stackrel{iid}{\sim}  \sum^{K_\omega}_{k=1} \pi_{\omega,k}\MVN(.|\umu_{\omega,k}, \Sigma_{\omega,k})$, $\umu_{\omega,k} \sim \MVN(.|\umu_{\omega,0}, \Sigma_{\omega,0})$, and  $\Sigma_{\omega,k} \sim \Invwish(.|\nu_0, \Psi_{0,\omega})$ constrained to $\sum^{K_\omega}_{k=1}  \pi_{\omega,k} \umu_{\omega,k} = \bzero$. For the cluster probability, we assume $\upi_{\omega} = (\pi_{\omega,1}, \cdots, \pi_{\omega, K_\omega})\trans \sim \Dir(\alpha_u/K_\omega, \cdots, \alpha_u/K_\omega)$. Additionally, we assume $\uX_i \sim  \sum^{K_x}_{k=1} \pi_{x,k} \MVN(.|\umu_{x,k}, \Sigma_{x,k})$, with $\umu_{x,k} \sim \MVN(.|\umu_{x,0}, \Sigma_{x,0})$, $\Sigma_{x,k} \sim \Invwish(.|\nu_{x,0}, \Psi_{x,0})$, where
$\upi_{x} = (\pi_{x,1}, \cdots, \pi_{x,K_x})\trans$ 
$\sim \Dir(\alpha_x/K_x, \cdots, \alpha_x/K_x)$. We use a very diffused prior $p(\ubeta_z) \propto 1$. Finally, the smoothing prior for $\ugamma \propto \exp\left\{-\frac{1}{2 \tau} \ugamma\trans\uP_{\gamma} \ugamma\right\}$ and $\tau \sim \IG(\alpha_0, \beta_0)$, where $\uP$ is the second order difference penalty matrix similar to that used by \cite{lang2004bayesian}. Note that $\tau$ represents a smoothing parameter, with smaller values of $\tau$ leading to stronger smoothing. The choice of a penalized prior on $\ugamma$ eliminates the need to choose the number basis $K$ precisely in the model since unnecessary basis functions will be heavily penalized and their coefficients driven to zero \citep{RWC2003,lang2004bayesian}. 

\subsection{Posterior} \label{sec:post}
We sample from the joint posterior distribution using a sequence of straightforward Gibbs steps with the full conditional posterior distribution completely defined. We describe the choice of prior hyper-parameters and starting values for the Markov Chain Monte Carlo (MCMC) in Section S3 and the form of the full conditional posterior distributions in Section S4 of the online supplementary material. Software ({\tt R code}) for implementing our method, a sample input data set, and complete documentation are available at {\tt https://github.com/rszoh/SoFR-ME}. With our Bayesian approach, all inferences about $\ugamma$ (hence $\ubeta(t)$) can be made simply based on the draws from the posterior distributions and no additional bootstrap approach is needed. Our Bayesian model also enables extracting subgroups that differ in term of their true unobserved (latent) $\uX$ function value, as the truncated Dirichlet process prior amounts to performing clustering \citealp{li2019tutorial}. However, we use our tDPM to allow for flexible modeling of the error term distribution, not for clustering.  

\subsection{Estimation of $\udelta(t)$} \label{sec:deltestim}
To this point, we have assumed that $\udelta(t)$ is known. In application, $\udelta(t)$ in unknown and must be estimated from the data. We propose a simple approach to estimating $\udelta(t)$ at each time point. Based on Equations~\ref{eq:modeq2} and ~\ref{eq:modeq7}, $E\{W_i(t)\} = E\{X_{i}(t)\}$ and $E\{M_i(t)\} = \udelta(t) E\{X_i(t)\}$. Therefore, $\udelta(t) = E\{M_i(t)\} / E\{W_i(t)\}$ 
provided $ E\{X_i(t)\} \neq 0$. So we estimate $\udelta(t)$ as
\be
\widehat{\udelta}(t) &=& \frac{ \sum^{n}_{i=1}M_i(t)}{ \sum^{n}_{i=1}W_{i}(t)} \label{eq:delta}
\ee
If we further assume that $\udelta(t)$ is a smooth function,  any of the local polynomial smoothing approaches available in R can be used to smooth the point-wise estimates obtained from Eq.~\ref{eq:delta}. 

\section{Simulation \& Results} \label{sec:simeval}
\subsection{Simulation Set-up}
For each $i =1,\cdots, n$ independently, we simulated $X_i(t)$ from a Gaussian process with $E\{X_i(t)\} = \{\sin(2\pi t) + 1.25\}/2$, $\var\{X_i(t)\} = \sigma^2_x $, and $\cov \{ X_i(t_l), X_i(t_j) \} = \rho_x \sigma^2_x $. We simulated $W_i(t) = X_i(t) + U_{i}(t)$, where $U(t)$ is a Gaussian process with $E\{U(t)\} = 0$, $\var\{U_i(t)\} = \sigma^2_u $, and $\cov\{U_i(t_l), U(t_j)\} = \rho_u \sigma^2_u$; and $M_i(t) = \udelta(t)X_{i}(t) + \omega_i(t)$, where $\omega_i(t)$ is a Gaussian process with $E\{\omega_i(t)\} = 0$, $\var\{\omega_i(t)\} = \sigma^2_{\omega} $, and $\cov\{\omega(t_l), \omega(t_j)  \} = \rho_\omega \sigma^2_{\omega}$. Finally, we simulated the response for each unit as $Y_i = \uZ_{i}\trans\ubeta_z + \int^1_{0} \ubeta(t)X_{i}(t)dt + \epsilon_i$, where $\epsilon_i \sim \uG$, for some distribution $\uG$. We simulated these data assuming $\ubeta_z = \bzero$.
In each simulation exercise, we assessed the performance of each estimator of $\ubeta(t)$ with the average bias squared ($\text{ABias}^2$), average sample variance (\text{Avar}), and the  mean squared integrated error (\text{MISE}):
\be
ABias^2 (\widehat{\beta}) &=& \frac{1}{n_{grid}}\sum^{n_{grid}}_{l=1}\left\{ \bar{\beta}(t_l) - \beta(t_l) \right\}^2,\\
Avar(\widehat{\beta}) &=& \frac{1}{n_{s} n_{grid} }\sum^{n_{r} }_{i=1}\sum^{n_{grid}}_{j=1}\left\{ \widehat{\beta}(t_j) - \bar{\beta}(t_j)\right\}^2 {, and }\\
MISE &=& ABias^2 + Avar,
\ee
where $n_{grid}$ represents the number of equally selected grid points between 0 and 1 and $\bar{\beta}(t_j) = \frac{1}{n_r}\sum^{n_{r}}_{j=1}\widehat{\beta}^{r}(t_j)$ represents the point-wise average over the $n_r$ replicates of $\beta(t_j)$ at a specific time point $t_j$.
We denote $\beta(t)$ estimates from our approach as $\widehat{\bbeta}_{BIV}$ (Bayesian approach using IV), $\widehat{\bbeta}_{IV,\delta}$ (scaling $\uM$ by $\udelta(t)$) and $\widehat{\bbeta}_{IV}$ (not scaling $\uM$) are both variants of our prior approach \citealp{tekwe2019instrumental}. \textcolor{black}{Note the approach proposed by Tekwe et. al.~\cite{tekwe2019instrumental} is a non-Bayesian approach that relies on the generalized method of moments (GMM) to estimates $\beta(t)$ with no model assumption;} $\widehat{\bbeta}_{ReF.W}$ denotes the estimated $\bbeta(t)$ from the function {\bf \tt{pfr}} in the R package refund~\cite{refundMan}, treating $\uW$ as the true functional covariate followed by a penalized B-splines option \textcolor{black}{using traditional functional data methodology}. We describe in detail the options we used for the {\bf \tt{pfr}} function in our simulation in the online supplementary material; \textcolor{black}{$\beta_{W}$ estimates $\beta(t)$ using Equation~\ref{eq:modeq3} assuming that $\uW$ are the true covariates in lieu of the unobserved covariates $\uX$}. 

Finally, we choose $\udelta(t) = \frac{\delta}{2}(1+\sin(2\pi t))+ \min(\delta, 0.005)$. For each replication, we choose $K = 15$. We performed $N = 500$ replications in each simulation we ran. The number of basis is set at $K = 15$, and $K_x = K_u = K_{\omega} = K_{\epsilon} = 3$. The choice of all prior hyper-parameters is discussed in section S3 of the online supplemental material. 

\subsection{Simulation Results}
We ran four simulations. In the first, we compared the performance of all competing approaches. We simulated data assuming $\sigma_x = 4.0 $, $\sigma_u = 4.0$, $\sigma_{e} = 1.0$,  $\sigma_{\omega} = 1.0$, and $\rho_x = \rho_u = \rho_{\omega} = 0.25$, with $n = 500$ and $1000$. We evaluated the functional covariate $X(t)$ at $t=50$ different but equidistant time points on the $(0, 1)$ interval. We simulated the error $\epsilon_i \stackrel{iid}{\sim} \Normal(0, \sigma^2_e)$ and $\sigma_e = 1$. Our approach ($\bbeta_{BIV}$) had the lowest MISE. $\bbeta_{IV}$ had the lowest bias ( Table~\ref{tab:tbl1}) and the approaches based on $\uW$ had the highest biases. Also, adjusting $\bbeta_{IV}$ for $\delta(t)$ introduced more bias in the estimate and ultimately destroyed the performance of $\bbeta_{BIV, \delta}$. Our Bayesian estimator performed better than its direct non-Bayesian counterpart. Clearly, the assumption of constant $\udelta(t)$ (i.e, $\udelta(t) = 1.0$) made in the non-Bayesian approach is violated. We also performed the same simulation assuming $\udelta(t) = 1.0$, and our Bayesian approach again outperformed the other approaches  (see Table S1 in the online supplementary material).

In the second simulation, we investigated the impact of different levels of the variances $\sigma_x$, $\sigma_u$, and $\sigma_{\omega}$ on $\bbeta(t)$. We evaluated variance values of 0.5, 1.0, 4.0, and 16. We changed only one variance term at a time and held the others constant at a value of 1.0 and $\sigma_x = 4.0$ when not varying. We simulated data assuming $\delta = 2$, $\sigma_{e} =1.0$, and $\rho_x = \rho_u = \rho_{\omega} = 0.25$, with $n = 500$. For all approaches, the estimated MISE increased with increasing $\sigma_u$ and $\sigma_{\omega}$ but decreased with increasing $\sigma_x$ (Table~\ref{tab:tabl2}). However, $\bbeta_{BIV}$ had the lowest MISE in all cases. All approaches tended to have higher bias with increasing $\sigma_u$ and $\sigma_{\omega}$ although the increase in the bias was much larger for the non-Bayesian $\bbeta_{IV}$ than the Bayesian $\bbeta_{BIV}$, especially for larger variances. Namely, $\bbeta_{IV}$ can get severely biased with large values of $\sigma_u$ and $\sigma_{\omega}$ and low values of $\sigma_x$ compared to its Bayesian counterpart. 

In our third simulation, we assessed the impact of varying levels of dependency in the error processes for $X(t)$, $U(t)$, and $\omega(t)$ on the estimate of $\ubeta(t)$. We evaluated correlation coefficient values for $\rho_x, \rho_u, and \rho_w$ of 0, 0.25, 0.50, and 0.75. We simulated data assuming $\sigma_x = 4.0 $, $\sigma_u = 4.0$, $\sigma_{e} = 1.0$, $\sigma_{\omega} = 1.0$, and $n = 500$. We changed one correlation parameter at a time and held the others constant at $0.25$. Overall, the Bayesian approach ($\bbeta_{BIV}$) outperformed its non-Bayesian direct competitor ($\bbeta_{IV}$) in terms of the MISE (Table~\ref{tab:tabl3}). Larger values of $\rho_u$ and $\rho_w$ did not affect the MISE for $\bbeta_{BIV}$ but they did increase the MISE for $\bbeta_{IV}$ substantially. Increasing values of $\rho_{x}$ tended to increase MISE values for both estimators. The combination of non-constant $\delta(t)$ (our second simulation) and large correlation values for $\rho_u$ and $\rho_w$ (this simulation) therefore impacts the performance of $\bbeta_{IV}$ significantly.

In our fourth simulation, We assessed the impact of varying the magnitude of the function on both our Bayesian estimator and its non-Bayesian counterpart. We simulated data for different values of the function $\udelta(t) = \frac{\delta}{2}(1+\sin(2\pi t))+ \min(\delta, 0.005)$ based on values of 0.5, 5, and 20 for $\delta$. Large values of $\delta$ lead to values of $\udelta(t)$ far from zero. We simulated data assuming $n = 500$, $\sigma_x = 4.0 $, $\sigma_u = 4.0$, $\sigma_{e} =1.0$, $\sigma_{\omega} = 1.0$, and $\rho_x = \rho_u = \rho_w = 0.25$. The MISE decreased with increasing $\delta$ for both approaches although the Bayesian approach had the lowest MISE in all cases (Table~\ref{tab:tabl4}). Both approaches had high MISE when $\udelta(t)$ was closer to zero. This is consistent with expectations for an instrumental variable. That is, $\udelta(t) = 0$ corresponds to a poor instrument, leading both approaches to produce unreliable estimates.     

Throughout our presentation, we have assumed that $\udelta(t)$ is known. In real applications, $\udelta(t)$ is unknown but can be estimated (section~\ref{sec:deltestim}). We simulated data estimating $\udelta(t)$ in lieu of the fixed values of true $\udelta(t)$ (Table S2 in the online supplemental material). The estimated MISE was higher when $\udelta(t)$ was estimated than when it was fixed. Estimates of $\udelta(t)$ were unstable at time points where $\uX(t)$ were small and close to zero (Figure S1 in the online supplementary material). Eliminating these occurrences from the simulation yield estimate of $\udelta(t)$ very closed to the truth (Figure S2 of the online supplementary material.) We also performed additional simulations assuming non-symmetric (Centered \text{Gamma} distribution and a mixture of two-normal) distributions of the error terms for the response $Y$ (Table S3 in the online supplementary material). Again, our Bayesian approach seems to perform much better than its non-Bayesian counterpart, suggesting that our approach adapts well to the non-normality of the response.

\section{Application}\label{sec:application}
\begin{figure}[htb]
\centering
\caption{Plot the analysis summary.
(a) Spaghetti plot of the $\log2$ physical intensity patterns with the average for all participants (in cyan,long-dashed line), for group 1 (in red - continuous line), for group 3 (in black - dashed line); (b) Spaghetti plot of the $\log2$ step counts patterns with the average for all participants (in cyan, long-dashed line), for group 1 (in red - continuous line), for group 3 (in black - dashed line); (c) Plot of the estimated functional parameter $\bbeta(t)$ based on the Bayesian approach $\bbeta_{BIV}$ and its non-Bayesian counterpart $\bbeta_{IV}$ along with the bound of a point-wise $90\%$ credible posterior intervals; (d) Estimate posterior mean of the latent covariate $\uX$ for the 3 latent physical activity pattern observed. }
 
\subfloat[]{\includegraphics[width = 2.5in]{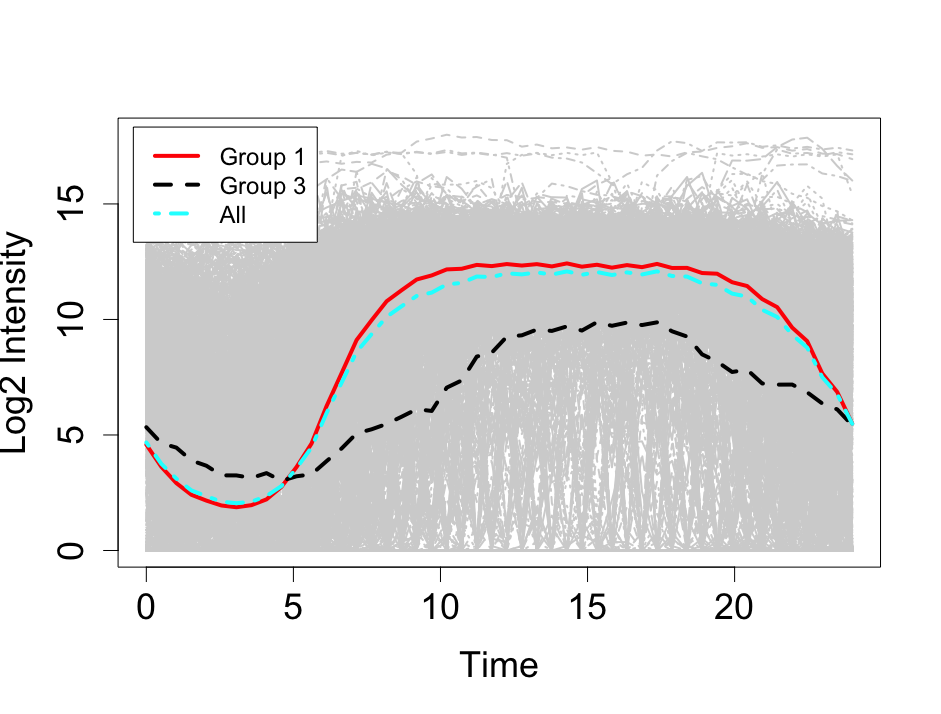}\label{fig:a}}
\subfloat[]{\includegraphics[width = 2.50in]{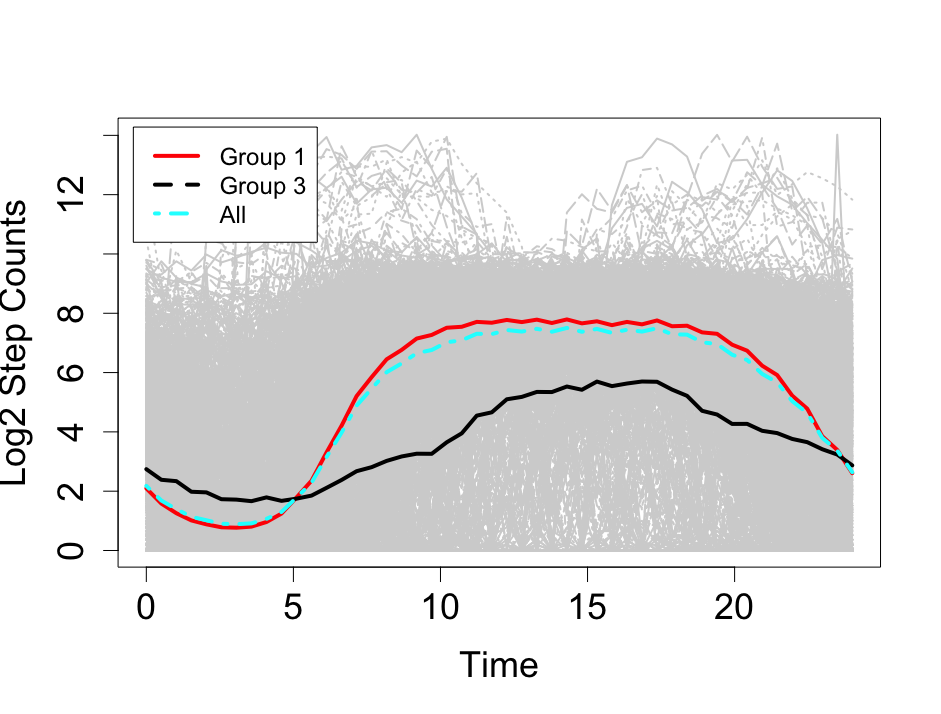}\label{fig:b}}\\
\subfloat[]{\includegraphics[width = 2.5in]{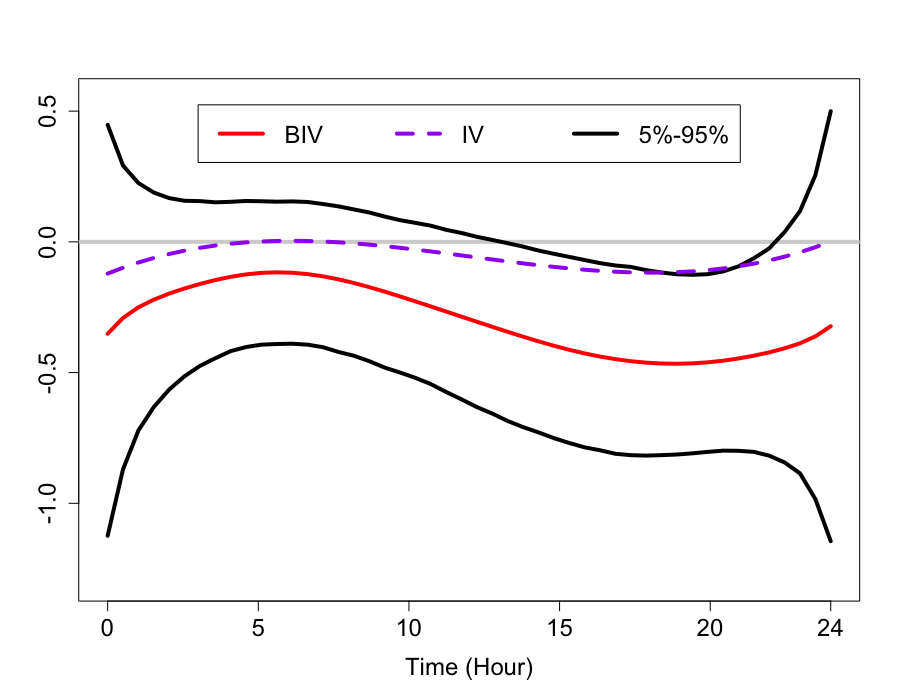}\label{fig:c}}
\subfloat[]{\includegraphics[width = 2.5in]{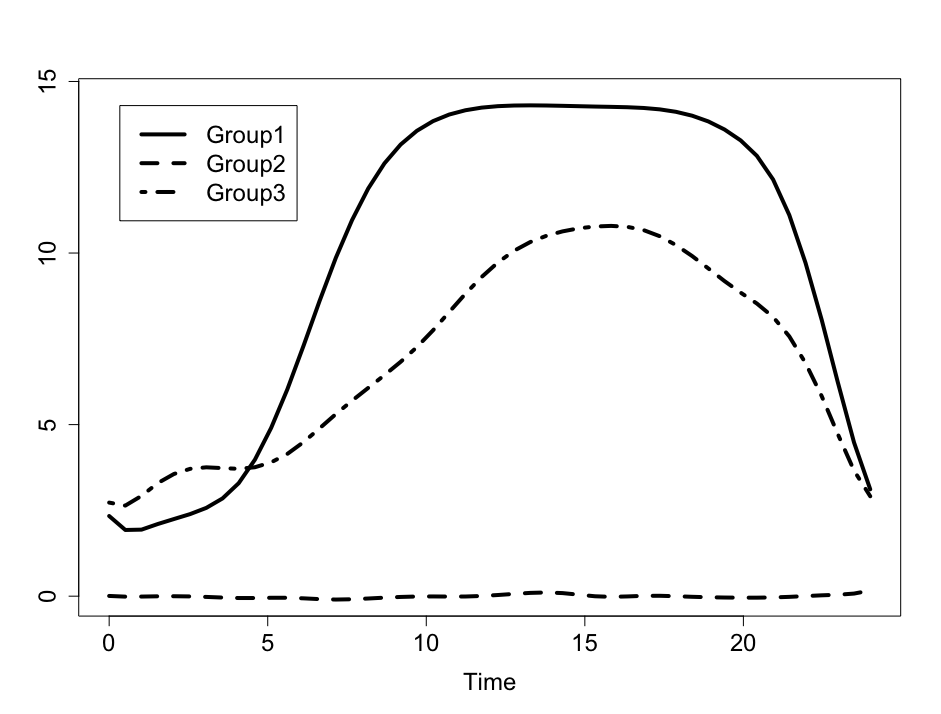}\label{fig:d}}
\label{fig:fig2}
\end{figure}

The motivating data for our method come from the National Health and Examination Survey (NHANES). The Centers for Disease Control and Prevention established the NHANES program in the 1960s to describe the health of Americans. This periodic cross-sectional survey involves a complex multi-stage cluster probability sampling scheme and participant sampling weights so that the resulting data represent the U.S. population as a whole. The data are based on in-person interviews and physical examinations. Approximately $5,000$ individuals from 15 geographic regions participate annually. NHANES data cover demographics, socioeconomic characteristics, diet, health behaviors, medical conditions, physiological examination results, and laboratory assessments of blood samples. Some participants in the 2005-6 cycles of the NHANES also wore accelerometer devices (ActiGraph AM-7164) on their hips for monitoring physical activity for at least four days during waking hours \citep{matthews2008amount,troiano2008physical}. Participants removed the devices during water activities and sleep. The devices recorded data on physical activity intensity and step count every 60 seconds while worn. In our analyses, we included $1,893$ NHANES participants who had accelerometer data for four or more days. Participants' median age was 43 years (range = 20\textendash 69 years), and 49\% were women.  
We considered true physical activity intensity as the true function-valued covariate, $X(t)$, the device-based observed measure of physical activity as $W(t)$, and the device-based measure of step count as the IV, $M(t)$. We assumed that step count is independent of measurement error conditional on true physical activity intensity. We also assumed that the association of step count with the device-based measure of physical activity intensity is only through its association with true physical activity intensity. The error free covariates included in our analyses were age, sex, and ethnicity. Following the NHANES analytic guidelines \citep{johnson2013national}, we applied sample weights to the data to account for the over sampling of racial groups. We grouped the device-based data into $30$-minute (half-hour) intervals and we averaged across the multiple days of wear time. We standardized the data by defining our baseline $t=0$ as the time when a participant began wearing the device, while $t=24$ represents the end of the day. Figures ~\ref{fig:a} and \ref{fig:b} show half-hour average intensity ($\log2$ transformed) and plots of half-hour average step counts ($\log2$ transformed),  and BMI values (Figure S3 of the online supplementary material). 

We applied a $\log2$ transformation on $\uM$ and $\uW$, and then estimated $\hat{\udelta}(t)$ (section 2.4 and Figure S4 in the online supplementary material). \rsz{We observe that $\hat{\udelta}(t)$ is not constant and ranges between 0.4 and 0.7 with high values during day time, suggesting that a time varying $\delta(t)$ maybe useful. We use the priors described in section~\ref{subsec:priors} and set $K = 20$, $K_{\epsilon} = K_u = K_x = K_{\omega} = 3$, and $\alpha_{\epsilon} = \alpha_{u} = \alpha_x = \alpha_\omega = 1$; $\mu_0 =0, \nu_0 = 0.1, \alpha_0 = 3, \sigma^{2}_0 =(\alpha_0 -1)\nu_0/(1+\nu_0) S^{2}_{y}$; where $S^{2}_{y}$ is the sample variance of the variance for the response $Y$; All the prior mean vectors are set to zero $\umu_{u,0} = \umu_{\omega,0} = \umu_{x,0} = \bzero$; $\Psi_{u,0} = \cov(\uW)$; $\Psi_{\omega,0}$ For the vector means and their covariance matrices, we choose $\nu_{u,0} = \nu_{\omega,0} = \nu_{x,0} = 5.0 (K_{\epsilon}+2)$ and $\Psi_{u,0} = \cov(\uW)$, $\Psi_{\omega,0} = \cov(\uM^{\star})$. Finding values for $\umu_{x,0}$, $\Sigma_{x,0}$ and $\Psi_{x,0}$ is a little more involve and we describe how to obtain them in details in Section S.3.2 of the online supplementary material.}

\textcolor{black}{
We note that the resulting posterior inference is mostly unchanged with small changes in the prior distribution choice we discussed above}. For the analysis, we ran our MCMC for $20,000$ iterations, discarded $5000$ as burn-in, and selected values of $3$ for thinning. We used the remaining MCMC draws for the final inference. We did not detect any obvious issues with all MCMC convergence diagnostics. Figures~\ref{fig:a} and ~\ref{fig:b} show the plots of $\log2$ transformed physical activity intensity ($\uW(t)$) and step count ($\uM(t)$) along with the mean curves of the three groups of physical activity obtained from the median posterior probability $\pi_{x,k}$ (k = 1,2,3) for each individual. 
Figure~\ref{fig:c} shows the estimated functional parameter $\ubeta(t)$ for both our Bayesian approach and its non-Bayesian direct competitor $\ubeta_{BIV}$,  and $\ubeta_{IV}$), respectively. The estimate of $\ubeta(t)$ for $\ubeta_{IV}$ is based on the regression residuals obtained after regressing $Y$ (BMI) on the error free covariates $\uZ$. The posterior mean of $\ubeta_{IV}$ is negative overall but seems to increase at decreasing physical activity windows (0 to 5hrs and around 20 to 24hrs) but decreases at increasing or high physical activity windows (5 to 20 hrs). We note that for our Bayesian estimate, the point-wise $90\%$ credible posterior intervals do not contain the zero around time widow 15hrs to 22hrs. This suggests that physical activity might have some association with lower BMI. The estimate based on $\ubeta_{IV}(t)$ closely resemble our estimate $\ubeta_{BIV}(t)$ but seems slightly attenuated towards the zero vertical line suggesting weaker physical activity effect on BMI compared to what are observed for our the Bayesian estimator.     
 
In addition to estimating the function parameter $\ubeta(t)$, it might be of interest to extract potential individual groups that differ in term of their true (latent) physical activity pattern. Our Bayesian semi-parametric model allows one to infer the potential grouping of the true latent physical activity pattern along with the group mean functions. Base on our posterior distribution of the group membership indicator, we estimate two main physical activity pattern along with their respective mean functions (see Figure~\ref{fig:d}). The size of these groups are: 1653 (Group 1), 0 (Group 2), and 240 (Group 3).  We estimate the posterior mean difference BMI between two similar individuals (race and age) in Group 1 and Group 3 based on their mean cluster physical activity as $\int^{1}_0 \beta(t)\{ \hat{\uX}_{\text{Group 1}} - \hat{\uX}_{\text{\text{Group 3} }}  \} dt$  which we estimated to be $-0.958\; (-1.560, -0.354)$ along with a $90\%$ posterior interval for the difference - which does not contain zero. This suggests that these two groups might be significantly different in term of their predicted BMI based on their respective group average physical activity. We also report the estimates for the error free covariates (see Table~\ref{tab:tabl5}) along with their $90\%$ posterior credible intervals. All the posterior credible interval don't contain zero except for the term gender.  
\section{Conclusion} \label{sec:conclu}
We proposed a fully Bayesian approach to estimate the functional parameter in SoFR with error free covariates and a functional covariate contaminated by measurement error. We used a potentially biased functional instrumental variable for model identification. In extensive simulations, our proposed Bayesian estimator had lower MISE than its non-Bayesian direct competitor.
Our approach does not require an unbiased IV. Instead, it allows the functional IV to have a time-varying biasing factor $\udelta(t)$ estimated from the data. As shown in our simulations, assuming this time-varying biasing factor $\udelta(t)$ is constant and equal to $1$ can lead to biased estimates.
Although ours is the first fully Bayesian approach to a measurement error for SoFR with an instrumental variable, it can be extended substantially. The assumptions that the error terms are independent of the unobserved covariate, error free covariates, and biasing function can be relaxed. Additionally, our MCMC, although fast, can be also be improved with a tailored variational Bayes approach. 


\begin{table}[ht]
\centering
 		\caption{Estimated $MISE$ based on $500$ replicates. Data are simulated assuming no effect for the error-free covariates.}
		\label{tab:tbl1}
\begin{tabular}{|rl|rrr|}
  \hline
 Method & n & $\text{ABias}^{2}$ & AVar & MISE \\ 
  \hline
\multirow{2}{*}{$\bbeta_{BIV}$} & 500  &   0.0140 & 0.0285 & 0.0425 \\ 
   & 1000  &   0.0014 & 0.0231 & 0.0245 \\ \hline
 	\multirow{2}{*}{$\bbeta_{IV,\delta}$} & 500   & 0.0441 & 0.3700 & 0.4141 \\ 
   & 1000   & 0.0234 & 0.4245 & 0.4479 \\  \hline
  \multirow{2}{*}{$\bbeta_{IV}$} & 500 &   0.0003 & 0.2238 & 0.2241 \\ 
    & 1000 &  0.0002 & 0.2784 & 0.2786 \\ \hline
 \multirow{2}{*}{$\bbeta_{IReF.W}$} & 500 &   0.2911 & 0.0110 & 0.3021 \\ 
   & 1000 &   0.2912 & 0.0064 & 0.2976 \\ \hline
  \multirow{2}{*}{$\bbeta_{W}$} & 500 &   0.2687 & 0.0583 & 0.3270 \\ 
 & 1000 &   0.2676 & 0.0435 & 0.3111 \\ 
   \hline
\end{tabular}
\end{table}

\begin{table}[ht]
\centering
\caption{ Results from simulation with different values of $\sigma_\omega,\; \sigma_u$, and $\sigma_x$}
\label{tab:tabl2}
\resizebox{\columnwidth}{!}{%
	\begin{tabular}{|r|l|rrr|c|rrr|c|rrr|}
\hline
Method & $\sigma_u$ & ABias$^{2}$ & AVar & MISE & $\sigma_w$ & ABias$^{2}$ & AVar & MISE & $\sigma_x$ & ABias$^{2}$ & AVar & MISE \\ 
\hline
\multirow{4}{*}{$\bbeta_{BIV}$} & 0.5  & 0.0119 & 0.0223 & 0.0342 & 0.5  & 0.0094 & 0.0229 & 0.0324 & 0.5  & 0.1394 & 0.1249 & 0.2643 \\    & 1  & 0.0100 & 0.0243 & 0.0343 & 1  & 0.0112 & 0.0237 & 0.0349 & 1  & 0.0689 & 0.1025 & 0.1713 \\ 
   & 4  & 0.0128 & 0.0274 & 0.0401 & 4  & 0.0124 & 0.0269 & 0.0393 & 4  & 0.0127 & 0.0289 & 0.0416 \\ 
   & 16  & 0.0527 & 0.0425 & 0.0953 & 16  & 0.0389 & 0.0407 & 0.0796 & 16  & 0.0018 & 0.0035 & 0.0052 \\ \hline 
	\multirow{4}{*}{$\bbeta_{IV}$} & 0.5  & 0.0017 & 0.0794 & 0.0811 & 0.5  & 0.0017 & 0.0559 & 0.0576 & 0.5  & 0.9346 & 0.9350 & 1.8696 \\ 
    & 1  & 0.0018 & 0.0885 & 0.0903 & 1  & 0.0021 & 0.0829 & 0.0849 & 1  & 0.8979 & 0.9712 & 1.8691 \\ 
    & 4  & 0.0020 & 0.1906 & 0.1925 & 4  & 0.0016 & 0.0948 & 0.0964 & 4  & 0.0040 & 0.2013 & 0.2053 \\ 
    & 16  & 0.1024 & 0.7473 & 0.8498 & 16  & 0.0017 & 0.5407 & 0.5424 & 16  & 0.0016 & 0.0082 & 0.0098 \\ 
   \hline
\end{tabular}
}
\end{table}

\begin{table}[ht]
\centering
\caption{Results from simulation with varying levels of dependency in error terms $\rho_u, \rho_\omega$, and $\rho_x$}
	\label{tab:tabl3}
	\resizebox{\columnwidth}{!}{%
	\begin{tabular}{|r|rrrr|rrrr|rrrr|}
		\hline
	 Method & $\rho_u$ & ABias$^{2}$ & AVar & MISE & $\rho_{\omega}$ & ABias$^{2}$ & AVar & MISE & $\rho_x$ & ABias$^{2}$& AVar & MISE \\ 
		\hline
		\multirow{3}{*}{$\bbeta_{BIV}$} & 0.0 & 0.0133 & 0.0285 & 0.0418 & 0.0 & 0.0131 & 0.0265 & 0.0396 & 0.0 & 0.0171 & 0.0268 & 0.0439 \\ 
    & 0.25  & 0.0125 & 0.0263 & 0.0388 & 0.25  & 0.0121 & 0.0273 & 0.0394 & 0.25  & 0.0125 & 0.0291 & 0.0415 \\ 
    & 0.50 & 0.0130 & 0.0257 & 0.0387 & 0.50  & 0.0127 & 0.0268 & 0.0395 & 0.50  & 0.0252 & 0.0348 & 0.0600 \\ 
    & 0.75  & 0.0186 & 0.0260 & 0.0446 & 0.75  & 0.0126 & 0.0263 & 0.0389 & 0.75  & 0.0561 & 0.0527 & 0.1088 \\  \hline
  	\multirow{3}{*}{$\bbeta_{IV}$} & 0.0  & 0.0022 & 0.0873 & 0.0895 & 0.0 & 0.0014 & 0.1421 & 0.1434 & 0.0 & 0.0956 & 0.1017 & 0.1974 \\ 
   & 0.25  & 0.0049 & 0.2071 & 0.2120 & 0.25 & 0.0024 & 0.1818 & 0.1842 & 0.25 & 0.0022 & 0.2030 & 0.2052 \\ 
    & 0.50  & 0.0044 & 0.2906 & 0.2950 & 0.50 & 0.0036 & 0.2573 & 0.2609 & 0.50  & 0.0029 & 0.2721 & 0.2751 \\ 
   & 0.75  & 0.0213 & 0.4055 & 0.4267 & 0.75 & 0.0053 & 0.2634 & 0.2687 & 0.75  & 0.0070 & 0.5593 & 0.5664 \\ 
   \hline
\end{tabular}
}
\end{table}

\begin{table}[ht]
\centering
\caption{Estimated MISE for different values of the function $\udelta(t)$ (varying $\delta$).} 
\label{tab:tabl4}
\begin{tabular}{|lr|rrr|}
  \hline
  Method & $\delta$ & ABias$^2$ & AVar & MISE \\ 
  \hline
	\multirow{3}{*}{$\bbeta_{BIV}$}&  0.5 & 0.0154 & 0.0320 & 0.0475 \\ 
   &    5 & 0.0134 & 0.0274 & 0.0408 \\ 
   &    20 & 0.0138 & 0.0246 & 0.0384 \\ \hline
 	\multirow{3}{*}{$\bbeta_{IV}$} &    0.5 & 0.0482 & 0.3681 & 0.4163 \\ 
    &    5 & 0.0015 & 0.1239 & 0.1254 \\ 
   &    20 & 0.0018 & 0.1067 & 0.1085 \\ 
   \hline
\end{tabular}
\end{table}


\begin{table}[ht]
	\centering
	\caption{Posterior summary estimates for the error free covariates using the Bayesian Instrumental variable approach. Age was median centered by subtracting the median age (43 years).}
	\label{tab:tabl5}
	\resizebox{\columnwidth}{!}{%
	\begin{tabular}{|rrrrrrll|}
		\hline
		& Non-His\_WHite & Black & Hispanic & Asian & Others & Gender(Male) & Age (Median-43) \\ 
		\hline
		 Mean& 31.65 & 31.77 & 32.23 & 33.04 & 30.03 & 0.06 & 0.08 \\ 
		5\% & 29.80 & 29.75 & 29.93 & 31.08 & 27.84 & -0.45 & 0.05 \\ 
		95\% & 33.33 & 33.67 & 34.41 & 34.82 & 32.10 & 0.58 & 0.10 \\ 
		\hline
	\end{tabular}
}
\end{table}

%

\clearpage
\section*{\textit{Acknowledgement}}
Zoh's research was supported by National Cancer Institute Supplemental under award U01-CA057030-29S1.
Tekwe's research was supported by National Cancer Institute Supplemental under award U01-CA057030-29S2 and .
This research was also supported in part by Lilly Endowment, Inc., through its support for the Indiana University Pervasive Technology Institute.

\noindent {\bf{Conflict of Interest}}
\noindent {\it{The authors have declared no conflict of interest.}}

\bibliographystyle{plainnat}
\bibliography{Carmen_FMIMIC,Carmen_GFLRIV,Carmen_FMIMIC_cdt}

\end{document}